\documentclass[12pt]{article}
\usepackage{isolatin1,natbib,graphicx}

\begin{document}

\section*{\Large \bf Closest Star Seen Orbiting the
Supermassive Black Hole at the Centre of the Milky Way\footnote{based on observations at the Very Large
    Telescope (VLT) of the European Observatory in Chile}}

\bigskip

\bigskip

{\bf R. Schödel, T. Ott, R. Genzel$^{*}$, R. Hofmann, M. Lehnert}
\\
Max-Planck-Institut f\"ur extraterrestrische Physik, Giessenbachstr., 85748 Garching, Germany
\\$^{*}$ also Dept. Dept.of Physics, University of California, Berkeley CA 94720, USA
\\
{\bf A. Eckart, N. Mouawad}
\\
I. Physikalisches Institut, Universit\"at zu K\"oln, Z\"ulpicher Stra\ss e 77, 50937 K\"oln, Germany
\\
{\bf T. Alexander} 
\\
The Weizmann Institute of Science, Faculty of Physics, PO Box 26,
Rehovot 76100, Israel
\\
{\bf M.J. Reid}
\\
Harvard-Smithsonian Center for Astrophysics MS42, 60 Garden St,
Cambridge, MA 02138, USA
\\
{\bf R. Lenzen, M. Hartung}
\\
Max-Planck-Institut für Astronomie, Königstuhl 17, 69117 Heidelberg, Germany
\\
{\bf F. Lacombe, D. Rouan, E. Gendron}
\\
Observatoire de Paris - Section de Meudon, 5, Place Jules Janssen,
92195 Meudon C\'edex, France
\\
{\bf G. Rousset}
\\
Office National d'Etudes et de Recherches Aerospatiales, BP 72,
92322 Chatillon cedex, France
\\
{\bf A.-M. Lagrange}
\\
Laboratoire d'Astrophysique, Observatoire de Grenoble, BP 53, F-38041
Grenoble C\'edex 9, France
\\
{\bf W. Brandner, N. Ageorges, C. Lidman, A.F.M. Moorwood,
  J. Spyromilio, N. Hubin}
\\
European Southern Observatory, Karl-Schwarzschild-Str.2, D-85748
Garching, Germany
\\
{\bf K.M. Menten}
\\
Max-Planck-Institut für Radioastronomie, Auf dem Hügel 69, D-53121
Bonn, Germany\\

\bigskip 

{\bf Measurements of stellar velocities$^{1-6}$ and variable X-ray
  emission$^{7}$ near the centre of the Milky Way have provided the strongest
  evidence so far that the dark mass concentrations seen in many galactic
  nuclei $^{8}$ are likely supermassive black holes, but have not yet excluded
  several alternative configurations. Here we report ten years of high
  resolution astrometric imaging that allow us to trace two thirds of the
  orbit of the star currently closest to the compact radio source and massive
  black hole candidate SgrA*. In particular, we
  have observed both peri- and apocentre passages. Our observations show that
  the star is on a bound, highly elliptical Keplerian orbit around SgrA*, with
  an orbital period of 15.2 years and a peri-centre distance of only 17 light
  hours. The orbital elements require an enclosed point mass of
  $3.7\pm1.5\times10^{6}$ solar masses.  The data exclude with high confidence
  that the central dark mass consists of a cluster of astrophysical objects or
  massive, degenerate fermions, and strongly constrain the central density
  structure.}
  
\bigskip

For the past ten years we have been carrying out high resolution near-IR
imaging and spectroscopy of the central few light years of our Milky Way for a
detailed study of the stellar dynamics in the vicinity of the compact radio
source SgrA*$^{1-2,4,6}$, \ the most likely counterpart of the putative black
hole$^{9,10}$. From a statistical analysis of the stellar proper motions
(velocities on the plane of the sky derived from multi-epoch imaging data) and
line of sight velocities (Doppler motions derived from spectral lines) we
deduced the presence of a mass of $\sim $2.6 to 3.3 million solar masses
(M$_{\odot }$) concentrated within 10 light days of SgrA*$^{1,2,4}$. To
further improve the sensitivity (by $\sim $20) and the angular
resolution/astrometric precision of our study (by $\sim $3), we have begun
this year to use the new CONICA/NAOS adaptive optics assisted
imager/spectrometer on the 8m UT4 (Yepun) of the ESO VLT$^{11-13}$. Figure 1
shows a diffraction limited (60milli-arcsec (mas) FWHM) K$_{s}$-band (2.1 $\mu
m$) image of the central 40'' of the Milky Way taken with CONICA/NAOS in May
2002. A key factor in constraining the mass distribution is the alignment of
the infrared images, where the stars are observed, with the astrometrically
accurate radio images, where SgrA* is observed. For this purpose we aligned
our CONICA/NAOS images with the astrometric grid using 7 SiO maser sources in
the field of view (circles in Fig.1) whose positions are known through
measurements with the VLA and the VLBA to accuracies of a few milli arcseconds
(Reid M.J. et al., 2002, in preparation). Having thus derived astrometric
infrared positions for 2002, we were then able to compute exact stellar
positions relative to SgrA* (in right ascension and declination) for all
epochs (including data taken with the SHARP camera at the ESO NTT) between
1992 and 2002. The resulting position of the radio source SgrA* on the
infrared image has a 1$\sigma $ uncertainty of $\pm $ 10mas, or about a factor
3 better than previously$^{14}$.  The new position of SgrA* is $\sim$50 mas E
of the position given in reference 14. In spring 2002 the orbiting star S2 had
approached SgrA* to within 10-20mas, thus giving the unique opportunity of
determining the mass a factor 10-20 times closer in than in previous work.

\bigskip

The first measurements of orbital accelerations for S2 and S1, the two stars
closest to SgrA*, were consistent with orbits bound to a $\sim$3 million solar
mass central object, but still allowed a wide range of possible orbital
parameters$^{5,6}$. Specifically, possible orbital periods for S2 ranged from
15 to 500 years$^{5}$. With our new data, we are now able to determine a unique
orbit for S2 from astrometric proper motions and provide strong constraints on
the mass distribution on distances below one light day.  Figure 2 shows the
measured 1992-2002 positions of S2 relative to SgrA*. In spring 2002 we
happened to catch the peri-centre passage of the star, at which point the
measured velocity exceeded 5000 km/s, about 8 times greater than 6 years
ago$^{1,2}$ when S2 was at apo-centre. The S2 data points trace two thirds of
a closed orbit and are robustly fit by a bound Keplerian orbit around a
central point mass located at the position of SgrA*. The parameters of the
best fitting orbit, along with their fit and astrometric errors are given in
Table 1. They were derived using the publicly available Binary Star Combined
Solution Package$^{15}$. For the nominal SgrA* position, the uncertainties of
the fit parameters are generally $\leq 10\%$. The additional uncertainty
introduced by the astrometric errors is of similar size. The semi-major axis
(a=5.5 light days) and orbital period (15.2 yrs) imply a mass of 3.7($\pm
$1.5)$\times $10$^{6}$ M$_{\odot }$ within the peri-centre radius of 124 AU,
or 17 light hours.  The peri-centre passage of S2 in April/May 2002 thus
probes the mass concentration at $\sim $2100 times the Schwarzschild radius of
a 3 million solar mass black hole.  The peri-centre distance radius of S2 is
70 times greater than the distance from the black hole where the star would be
disrupted by tidal forces (about 16 light minutes for a $\sim $15M$_{\odot }$,
7R$_{\odot }$ star like S2$^{2}$).  Since tidal energy deposition falls faster
than the sixth power of the ratio of tidal radius to orbital radius, tidal
effects near the perinigricon of S2 are expected to be negligible, consistent
with its lack of infrared
variability.\\

\bigskip 

The remarkable consequence of the orbital technique is that the mass can be
determined from a single stellar orbit, in comparison to the statistical
techniques that use several tens to hundreds of stellar velocities at 10 to
300 light days from SgrA* (Fig.3).  In addition, the orbital technique
requires fewer assumptions than the other estimates (e.g.  equilibrium and
isotropy of orbits), and thus is less vulnerable to systematic effects.

\bigskip

The Galactic centre mass distribution resulting from all available data is
well fit by the combination of a $2.6\pm 0.2\times 10^{6}$M$_{\odot }$ point
mass (the supermassive black hole), plus the visible stellar cluster of core
radius 0.34 pc, an outer power-law density distribution with exponent $\alpha
$=1.8 and central density $3.9\times 10^{6}$M$_{\odot }$pc$^{-3}$ (Figure 3).
If the central point mass is replaced by a Plummer mass distribution, which is
the most compact one expected realistically (with a power-law index of
$\alpha=$5), in order to mimic the flatness of the observed mass distribution
over 3 order of magnitude in radius$^{4}$), its central density would have to
exceed 10$^{17}$M$_{\odot }$pc$^{-3}$, more than 4 orders of magnitude greater
than previous estimates$^{4,5,6}$. Such a Plummer distribution would be
appropriate if the dark mass consisted of a dark cluster of low mass stars,
neutron stars, or stellar black holes. The maximum lifetime of such a cluster
mass against collapse (to a black hole) or evaporation would be less than a
few 10$^{5}$ years$^{16}$, clearly a highly implausible configuration.
Further, theoretical simluations of very dense, core collapsed clusters
predict much shallower, near isothermal density distributions ($\alpha \sim
2$, see discussion in reference 2). We conclude that such a dark cluster\ 
model can now be safely rejected. Our new data also robustly exclude one of
two remaining, 'dark particle matter' models as alternatives to a supermassive
black hole, namely a ball of heavy (10-17 keV/c$^{2}$) fermions (sterile
neutrinos, gravitinos or axinos) held up by degeneracy pressure$^{17,18}$ ,
which in principle could account for the entire range of dark mass
concentrations in galactic nuclei with a single physical model. Because of the
finite size ($\sim $0.9\hbox{$^{\prime\prime}$} diameter) of a
non-relativistic, $3\times 10^{6}$M$_{\odot }$ ball of $\sim $16~keV fermions,
the maximum (escape) velocity is about 1700 km/s and the shortest possible
orbital period for S2 in such a fermion ball model would be about 37
years$^{18}$, clearly inconsistent with the orbit of S2. The enclosed mass at
perinigricon would require a neutrino mass of $>$50~keV, a value which can
safely be excluded for neutrino ball models trying to explain the entire range
of observed masses in galactic nuclei$^{17,18}$. The only dark particle matter
explanation that cannot be ruled out by the present data is a ball of bosons,
as such a configuration would have a radius only several times greater than
the Schwarzschild radius of a black hole$^{16,19}$ However, it would be very
hard to understand how the bosons first manage to reach such a high
concentration, and then avoid forming a black hole by baryonic
accretion$^{16,19}$.The data on the Galactic centre thus show that the central
mass distribution is remarkably well described by the potential of a point
mass over 3 orders in magnitude in spatial scale, from 0.8 light days to 2
light years. The contribution of the extended stellar cluster around SgrA* to
the total mass cannot be more than a few hundred solar masses within the
peri-centre distance of the orbit of S2.

\bigskip

In this letter we have presented the first step in a new phase of
near-infrared observations of the immediate surroundings of the central dark
mass in the centre of the Milky Way. The observation of orbits of stars
surrounding the central dark object offers a clean new way of constraining its
mass distribution and testing the supermassive black hole model with the
simple assumption of Keplerian orbits.  Within the next years we hope to
observe the accelerations and orbits of several faint stars near SgrA* that
have become observable with the increased resolution and sensibility of the
CONICA/NAOS camera/AO system at the VLT.  Even more detailed observations of
the SgrA* environment will become possible with infrared interferometry at the
Large Binocular Telescope, the ESO VLTI and the Keck interferometer, which
will provide a few to 10 mas (a few light hours) resolution and offer exciting
prospects for the exploration of relativistic motions at 10-100 Schwarzschild
radii from the central black hole$^{20}$.

\bigskip {\bf Acknowledgments} {\small We thank the teams who developed and
  constructed the near-infrared camera CONICA and the adaptive optics system
  NAOS. We are grateful to all the instrument scientists and ESO staff
  involved in the commissioning of CONICA/NAOS for generous observations of
  the Galactic Center. We thank C. H. Townes and J. Kormendy for valuable
  comments. We thank D. Gudehus for his friendly assistance with the
  Binary-Star Combined Solution Program.}

\newpage
\begin{table}
\begin{tabular}{cccc}
Parameter & Value &  Formal Error & Astrom. Error\\
\multicolumn{1}{l}{Black hole mass [10$^{6}\times$M$_{\odot}$]} & 3.7 & 1.0 & 1.1  \\
\multicolumn{1}{l}{Period [years]} & 15.2 & \multicolumn{1}{c}{0.6} & 0.8 \\ 
\multicolumn{1}{l}{Time of peri-centre passage [year]} & 2002.30 &
\multicolumn{1}{c}{0.01} & 0.05 \\ 
\multicolumn{1}{l}{Eccentricity} & 0.87 & \multicolumn{1}{c}{0.01} & 0.03\\ 
\multicolumn{1}{l}{Angle of line of nodes [degrees]} & 36 & \multicolumn{1}{c}{
5} & 8\\ 
\multicolumn{1}{l}{Inclination [degrees]} & $\pm$46 & \multicolumn{1}{c}{3} & 3 \\ 
\multicolumn{1}{l}{Angle of node to peri-centre [degrees]} & 250 & 
\multicolumn{1}{c}{4} & 3\\ 
\multicolumn{1}{l}{Semi-major axis [mpc]} & 4.62 & \multicolumn{1}{c}{0.39} & 0.43\\ 
\multicolumn{1}{l}{Separation of peri-centre [mpc]} & 0.60 &
\multicolumn{1}{c}{0.07} & 0.15\\
\end{tabular}
\caption{Derived orbital parameters for S2, their 1$\sigma$ errors resulting
  from the orbital fit and the errors due to the 10 milliarcsecond astrometric
  uncertainty. See the caption of Figure~2 for a description of the angles and
  of the errors..}
\end{table}
\bigskip
\vspace*{10cm}
\newpage

{\bf Figure 1: K$_{s}$-band image of the centre of the Milky Way.}\\
Left: Diffraction limited (60 mas FWHM) K$_{s}$-band (2.1$\mu $m) image of the
central $\sim 40\hbox{$^{\prime\prime}$}$ of the Milky Way, obtained with the
CONICA/NAOS adaptive optics imager on UT4 (Yepun) of the VLT on May 3rd, 2002.
North is up and East to the left, scales are for an assumed distance of 8 kpc
(26,000 light years)$^{4,21}$. The unique infrared wavefront sensor was used
to close the loop of the adaptive optics system on the bright supergiant IRS
7, $\sim $6\hbox{$^{\prime\prime}$} north of SgrA*. The Strehl ratio is 
$>$40\%. The radio positions of 7 SiO maser stars (open circles) were used to
align the infrared image with the radio astrometry frame (Reid et al. 2002, in
prep.).  The SiO masers originate in the central $\sim 1$ mas of the
circumstellar envelopes of infrared bright, red giants/supergiants. The
radio-to-infrared registration is accurate to $\pm 10$ mas (including the
effect of variation of the point spread function across the field), a factor 3
improvement over reference 14.  There we could only use two SiO sources,
giving the centre position, rotation angle and a single pixel scale for the
infrared images. Our new analysis allows solving, in addition, for second
order imaging terms (small for CONICA/NAOS, but significant for the earlier
SHARP/NTT data).  Right: The central $\sim 2\hbox{$^{\prime\prime}$}$ region
(rectangle in left inset) around the compact radio source SgrA* (cross). This
image is a sum of images taken in May 2002 and has been deconvolved with a
linear Wiener filter method to remove the seeing halos. The ring structures
around the brighter stars are artefacts of the linear deconvolution algorithm
that arise because information on the point spread function in Fourier space
is not known up to infinite frequencies.  Several of the stars near SgrA* are
marked, including the presently closest star (S2).

\bigskip

{\bf Figure 2: Orbit of S2 around SgrA*.}\\
Orbit of S2, relative to the position of SgrA* (large cross and circle,
denoting the $\pm 10$ mas uncertainties of the infrared-radio astrometry).
The filled small circles (with 1$\sigma $ errors) between 1992 and 2001, and
at 2002.50, denote the results of our speckle imaging with the SHARP camera on
the ESO NTT$^{4,6}$. The five open rectangles are the CONICA/NAOS data points
on 2002.25, 2002.33, 2002.40, 2002.58 and 2002.66. The projection of the best
fitting Kepler orbit is shown as a thick continuous curve, with main
parameters listed adjacent to the orbit (see also Table 1). For determination
of the orbital elements (Table~1) we used the publicly available Binary Star
Combined Solution Package$^{15}$, which computes best fit orbits from
position-time series (including their errors). Uncertainties in the derived
parameters are determined from a covariance matrix analysis. The additional
uncertainty introduced by the astrometric errors is of similar size and was
estimated by letting the position of SgrA* vary randomly within its 1$\sigma$
astrometric uncertainty limits and examining the change in the resulting
orbital parameters. Since we do not know the line-of-sight velocity of the
star, the sign of the inclination given in Table~1 is undetermined. The angle
of the line of nodes (36 deg) is counted E of N, with N up and E to the left.
The angle from node to peri-centre is counted from the node in the NE
quadrant in the direction of the motion of S2.

\bigskip

{\bf Figure 3: Mass distribution in the Galactic Centre.}\\
Mass distribution in the Galactic centre (for an 8 kpc distance$^{21}$). The
filled circle denotes the mass derived from the orbit of S2. The error bar
combines the orbital fit and astrometry errors (Table 1). Filled triangles
denote Leonard-Merritt projected mass estimators from a new NTT proper motion
data set by Ott et al. (2002, in preparation), separating late and early type
stars, and correcting for the volume bias in those mass estimators by scaling
with correction factors (0.88-0.95) determined from Monte Carlo modeling of
theoretical clusters$^{4}$. An open rectangle
denotes the Bahcall-Tremaine mass estimate obtained from Keck proper motions$%
^{3}$. Light filled rectangles are mass estimates from a parameterized
Jeans-equation model from reference 4, including anisotropy and
differentiating between late and early type stars. Open circles are mass
estimates from a parameterized Jeans-equation model of the radial velocities
of late type stars, assuming isotropy$^{4}$. Open rectangles denote mass
estimates from a non-parametric, maximum likelihood model, assuming isotropy
and combining late and early type stars$^{22}$. The different statistical
estimates (in part using the same or similar data) agree within their
uncertainties but the variations show the sensitivity to the input
assumptions. In contrast, the new orbital technique for S2 is much simpler
and less affected by the assumptions. The continuous curve is the overall \
best fit model to all data. It is a sum of a $2.6(\pm 0.2)\times 10^{6}$ M$%
_{\odot }$ point mass, plus a stellar cluster of central density $3.9\times
10^{6}$ M$_{\odot }$/pc$^{3}$, core radius 0.34 pc and power-law index $%
\alpha =1.8$ . The long dash-short dash curve shows the same stellar cluster
separately, but for an infinitely small core (i.e. a 'cusp'). The thick dashed
curve is the sum of the visible cluster, plus a Plummer model of a
hypothetical concentrated ($\alpha =5)$, very compact (R$_{o}=$0.00019 pc)
dark cluster of central density $1\times 10^{17}$M$_{\odot }$pc$^{-3}$.

\newpage

\thispagestyle{empty}
\begin{figure}
\includegraphics[width=\textwidth]{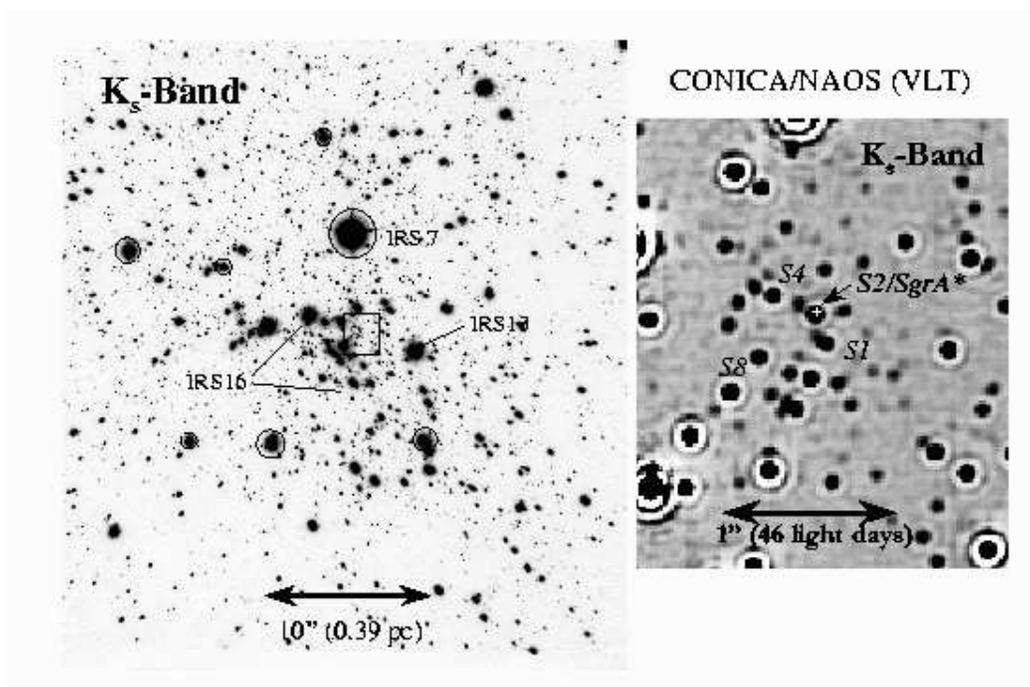}
  \caption{Rainer Sch\"odel, S08267}
\end{figure}

\newpage

\thispagestyle{empty}
\begin{figure}
  \includegraphics[width=\textwidth]{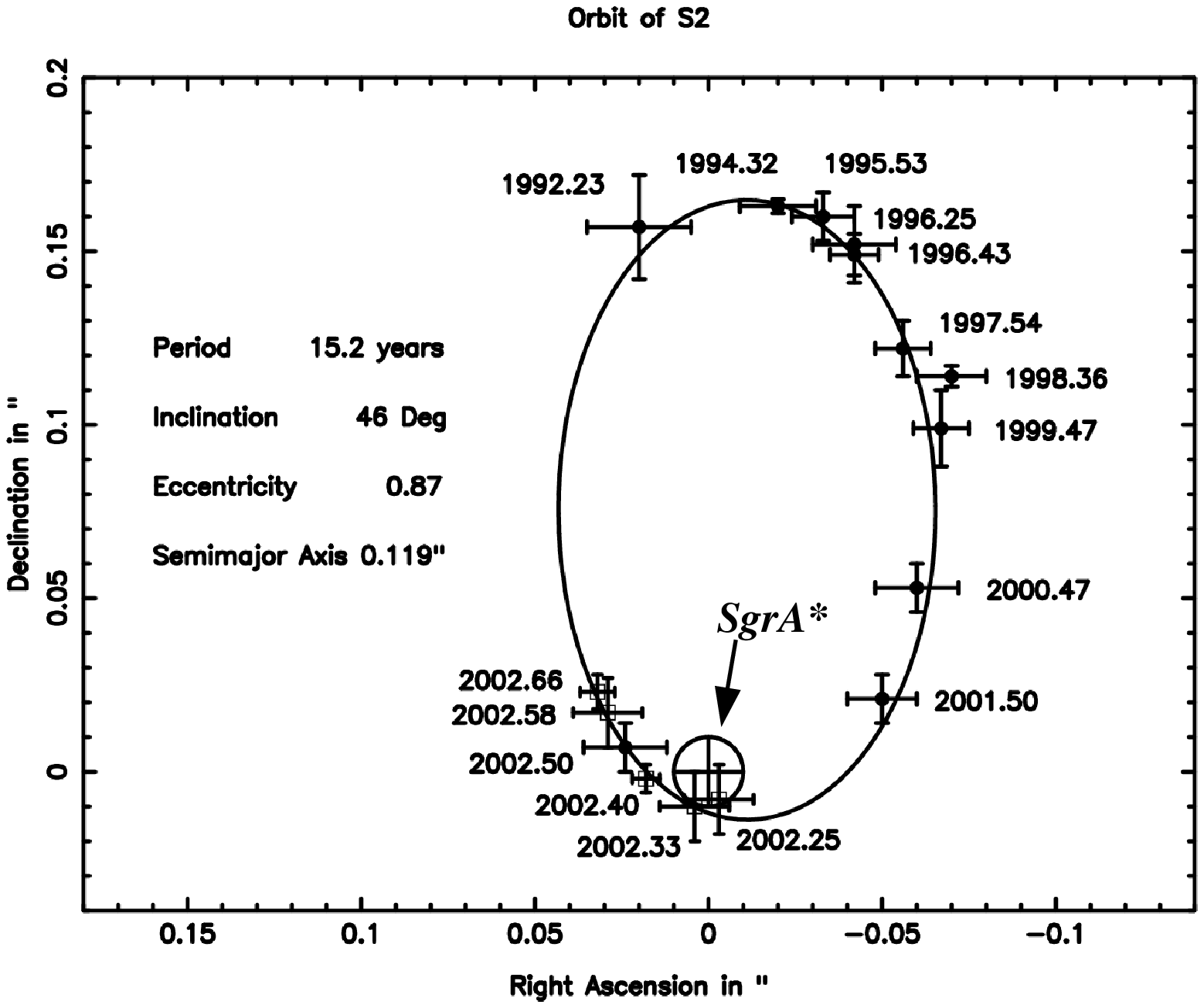}
  \caption{Rainer Sch\"odel, S08267}
\end{figure}

\newpage

\thispagestyle{empty}
\begin{figure}
  \includegraphics[width=\textwidth]{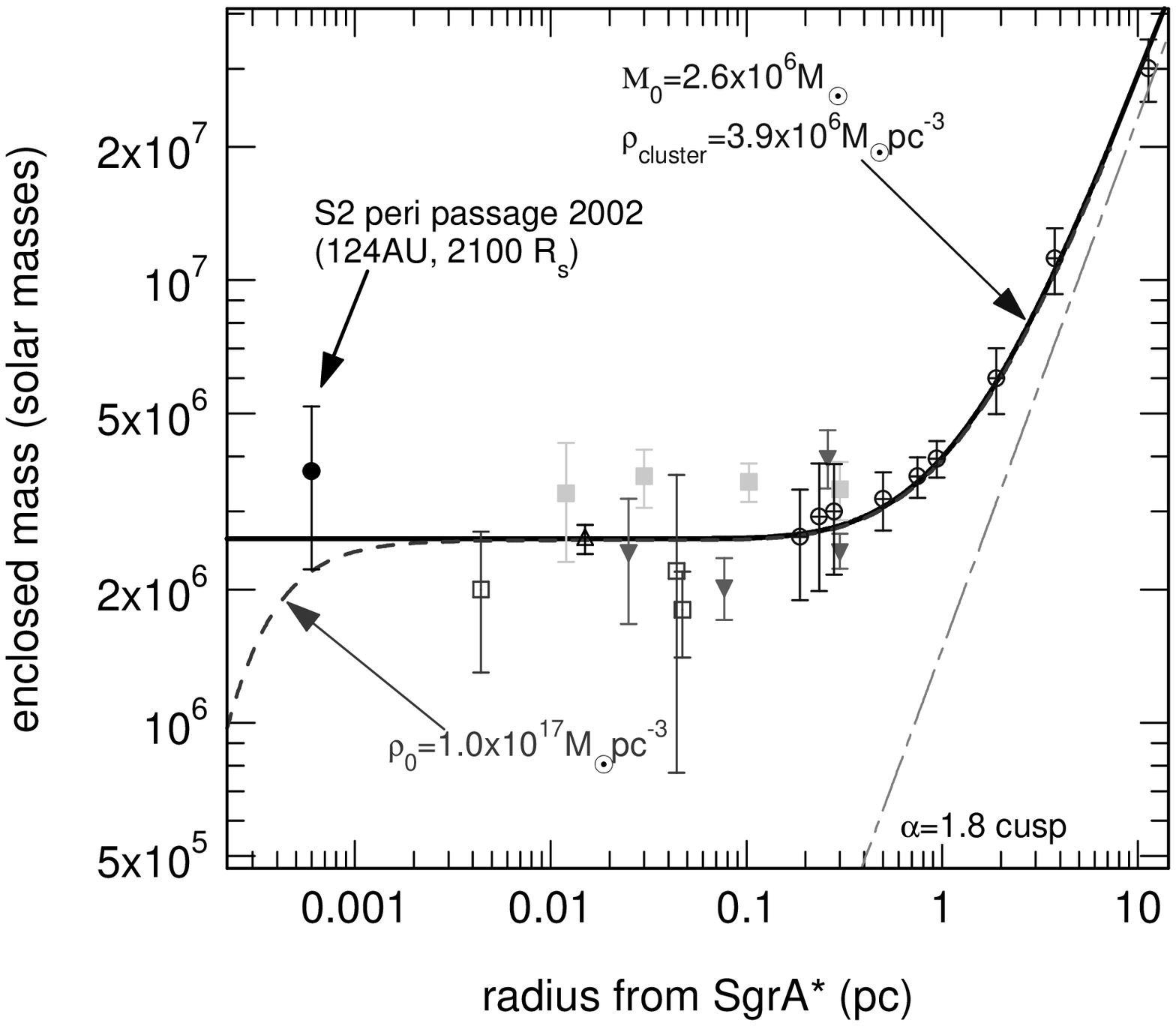}
\caption{Rainer Sch\"odel, S08267}
\end{figure}

\end{document}